\documentclass[allclo]{FBSart}
\usepackage{amsfonts,graphicx}
\usepackage{amssymb}

\def\be{\begin{eqnarray} &&}

\def\cal{\mathcal}
\def\ee{\end{eqnarray}}

\title{Relativistic Hamiltonian Dynamics and Few-Nucleon Systems}
\author{F. A. Baroncini\instnr{1}, E. Pace\instnr{1,2} and G. Salm\`e\instnr{3}}
\instlist{Dipartimento di Fisica, Universit\`{a} di Roma "Tor Vergata", 
 Via della Ricerca
Scientifica 1, I-00133 Roma, Italy \and
 INFN, Sezione di Tor Vergata, Via della Ricerca
Scientifica 1, I-00133 Roma, Italy\and
 INFN, Sezione di Roma, P.le A.
 Moro 2, I-00185 Roma, Italy}

\runningtitle{Relativistic Hamiltonian Dynamics and  Few-Nucleon Systems}
\runningauthor{F. A. Baroncini et al}

\begin{document}

\maketitle
\begin{abstract}
We present a preliminary calculation of the electromagnetic form factors 
of $^3$He and $^3$H, performed within  the Light-Front
Hamiltonian 
Dynamics. Relativistic effects show their relevance even at the static limit,
increasing at higher values of momentum transfer, as expected. 
\end{abstract}

\section{Introduction}

{The \em standard model} of Few-Nucleon Systems, where Nucleon and
pion degrees of freedom are taken into account, is already at a very
sophisticated stage, and many efforts are presently carried on to 
 retain all the {\em general principles} of a theory with a fixed
number of constituents. In particular,  to satisfy
Poincar\'e covariance
 appears a quite reachable goal
and at the same time a very compelling requirement in view of the forthcoming
measurements of electromagnetic (em) form factors (ff's) for A=3,4, in the region
of few GeV's \cite{Makis}. In order to  extract  unambiguous 
signatures of effects beyond 
the  {\em standard model} of Few-Nucleon Systems one should develop a fully
field-theoretical approach,  based on the Bethe-Salpeter
equation,  but the difficulties are well-known, and therefore one has to
consider 
 alternative approaches, like  3-D reductions 
  (see, e.g., \cite{Gross} for A=2
 and \cite{Stadler} 
for the work in progress for A=3) or the Relativistic Hamiltonian
 Dynamics (RHD) framework, suggested by Dirac in a seminal paper \cite{Dirac}.
 
 Our aim is  to construct, within the Light-Front form of  RHD, a relativistic 
 approach for Few-Nucleon
System that i) retains  the whole {\em successful phenomenology} 
already developed and 
ii)  includes,  {\em in a non perturbative way}, the relativistic
features requested  by Poincar\'e covariance. In order to have a strong and
immediate comparison with the experiments we have focused our efforts on the
development of Poincar\'e covariant calculations of the electromagnetic ff's,
extending our approach from the 
Deuteron \cite{LPS2K} to  the Trinucleon. 
The adopted Bakamjian-Thomas (BT) procedure (see, e.g., \cite{KP91}) allows us to  exploit 
 realistic wave functions for Few-Nucleon Systems
 (for A=3, see, e.g., \cite{KVR94})  in order to  evaluate 
 matrix elements of a 
Poincar\'e covariant current operator \cite{LPS98}  for an interacting 
system.   The relativistic effects
imposed by Poincar\'e covariance materialize in the relativistic kinematics and in
the presence of the so-called Melosh rotations (see, e.g., \cite{KP91}), that allows one to
use the standard Clebsh-Gordan machinery to obtain many-nucleon wave
functions with the correct angular coupling.
 
\section{Formalism} 
As well-known \cite{KP91}, Light-Front (LF) RHD has some appealing features, like the
largest number of  kinematical Poincar\'e generators (given the symmetry of the
{\em  initial} hypersurface $x^+=0$) and the simplest procedure
for separating out the center of mass  motion from the intrinsic one, in strict
analogy with the nonrelativistic procedure. Moreover, it  shares with  the other two RHD's
(Instant and Point forms)
the rigorous fulfillment of the Poincar\'e covariance, for a system with a fixed number of
constituents. 
In some sense, RHD's fall between non-relativistic quantum mechanics
and the local, relativistic field theory.

For an interacting system, an em 
 current operator, $J^\mu$,
that fulfills  the extended Poincar\'e  covariance (i.e.
including parity and time 
reversal) and  
Hermiticity, can be constructed by  a suitable auxiliary operator,
$j^\mu$, that fulfills rotational covariance around the $z$-axis in 
a Breit frame (${\bf P}_f+{\bf P}_i=0$), where the $\perp$ component of the momentum transfer is vanishing
(${\bf q}_\perp=0$) \cite{LPS98}. Note that such a frame is  {\em different} from the Drell-Yan one, where
$q^+ = 0$. In general \cite{LPS98},
 the matrix elements  $\langle P_f|J^\mu| P_i \rangle$, still acting on
 internal variables,   are directly given 
by the  matrix elements of the auxiliary operator $j^\mu$, evaluated in the
chosen Breit
frame. A minimal Ansatz for  
a {\em many-body} auxiliary operator is built from i)
the free current (a one-body operator) and ii) the $\perp$ component of the 
angular momentum operator
$\vec S$ (a many-body operator in LF)
 as follows
 \be   
j^{\mu}_{fi}(q\hat{e}_z)={1 \over 2 }~\left [{\cal{J}}^{\mu}_{fi} (q\hat{e}_z) +
L^{\mu}_{\nu}[r_x(-\pi)]~e^{\imath \pi S_x}~{\cal{J}}^{\nu}_{if}(q\hat{e}_z)^*
~e^{-\imath \pi S_x} \right ] 
\label{curlf}\ee
with
 $ 
{\cal{J}}^{\mu}_{fi}(q\hat{e}_z) = 
\Pi_f ~J_{free}^{\mu}(0) ~\Pi_i    
$,
 $\Pi \equiv$ projector onto the
  states 
of the  (initial or final) system and 
$\vec S \equiv$ the LF-spin operator of the system as whole: it acts on the
"internal" space and is unitarily related to the standard angular momentum
operator through the Melosh operators. Let us remind that 
$ J_{free}^{\mu}(0)$ is the proper sum over A=2,3 free Nucleon current given by 
 $J^\mu_{N}= -F_{2}[(p^{\prime
\mu}-p^\mu)^2](p^\mu+p^{\prime
\mu})/2M +\gamma^\mu (F_{1}[(p^{\prime
\mu}-p^\mu)^2]+F_{2}[(p^{\prime
\mu}-p^\mu)^2])$ , with $F_{1(2)}[(p^{\prime
\mu}-p^\mu)^2]$  the Dirac (Pauli) Nucleon
ff.
In our   Breit frame,  charge normalization and current conservation 
(for $M_f=M_i$) can be fulfilled by imposing 
 $ {\cal{J}}^{-}(q\hat{e}_z)={\cal{J}}^{+}(q\hat{e}_z)$\cite{LPS98,LPS2K}.

For evaluating matrix elements of $j^\mu(x)$, eigenstates of the 
 interacting system are needed. To this end one can use the "non relativistic
solutions", but with Melosh Rotations in the angular part, if the interaction
$V\equiv M_{int} -M_0$ (where $M_{int} (M_0)$ is the mass operator of the interacting 
(free) system) can be embedded in a BT
framework.
The BT construction for obtaining interacting Poincar\'e generators suggests a necessary (not sufficient) 
condition \cite{KP91} on the interaction: $V$
 must depend upon intrinsic variables  combined
in scalar products, i.e.
$ [\vec{\cal B}_{LF}, V ] = [\vec S_{0}, V ] = 
[P_{\perp}, V ] = [P^+, V ]= 0
$,
where $\vec{\cal B}_{LF}$ are the LF boosts, $\vec S_{0}\equiv$  the
angular momentum operator for the non interacting case (since
$S^2_{0}=S^2_{int}$ and $S_{0,z}=S_{int,z}$, the eigenvalues of $S^2_{0}$
and $S_{0,z}$ can label the eigenstates of the interacting system). The non
relativistic interaction fulfills the above requirements.

 \section{EM observables for A=2 and A=3 nuclei}

First, let us briefly review our results for the Deuteron, and then we present
preliminary calculations for the trinucleon case.

Magnetic and quadrupole moments of the Deuteron (see Table 1), as well as 
 the ff's, $A(Q^2)$ and $B(Q^2)$ (see Fig. 1),
and the tensor polarization $T_{20}(Q^2)$ have been calculated \cite{LPS2K},
showing the great relevance of the Poincar\'e covariance, even at low $Q^2$. 
From our results, one could argue that the role of MEC and  pair 
contributions should shrink, but clearly a direct evaluation of two-body
contributions is mandatory for a  closer comparison with experiments.
\begin{table} 
 \caption{Magnetic moment (in nuclear magnetons) and 
quadrupole moment
(in $fm^2$) for the
Deuteron \cite{LPS2K};  $P_D$ is the $D$-state percentage. The corresponding exp. values are:
 $\mu _{exp}=0.857406(1)$ and 
 $Q _{exp}= 0.2859(3)$}
 \begin{center}
\begin{tabular}{|c||c|c|c|c|c|}
\hline
Interaction & $P_D$   & $\mu _D^{NR}$ &
$\mu^{LFD}_D$  & $Q_D^{NR}$   & $Q^{LFD}_D$ \\
\hline  
CD-Bonn  & 4.83 &  0.8523  & 0.8670 & 0.2696 & 0.2729 \\
Nijm1    & 5.66 &  0.8475  & 0.8622 & 0.2719 & 0.2758 \\
RSC93    & 5.70 &  0.8473  & 0.8637 & 0.2703 & 0.2750 \\ 
Av18     & 5.76 &  0.8470  & 0.8635 & 0.2696 & 0.2744 \\
\hline 
\end{tabular}
\end{center} 
\end{table}
 \begin{figure}[t] 
\parbox{6.4cm}{\includegraphics[width=6.4cm] {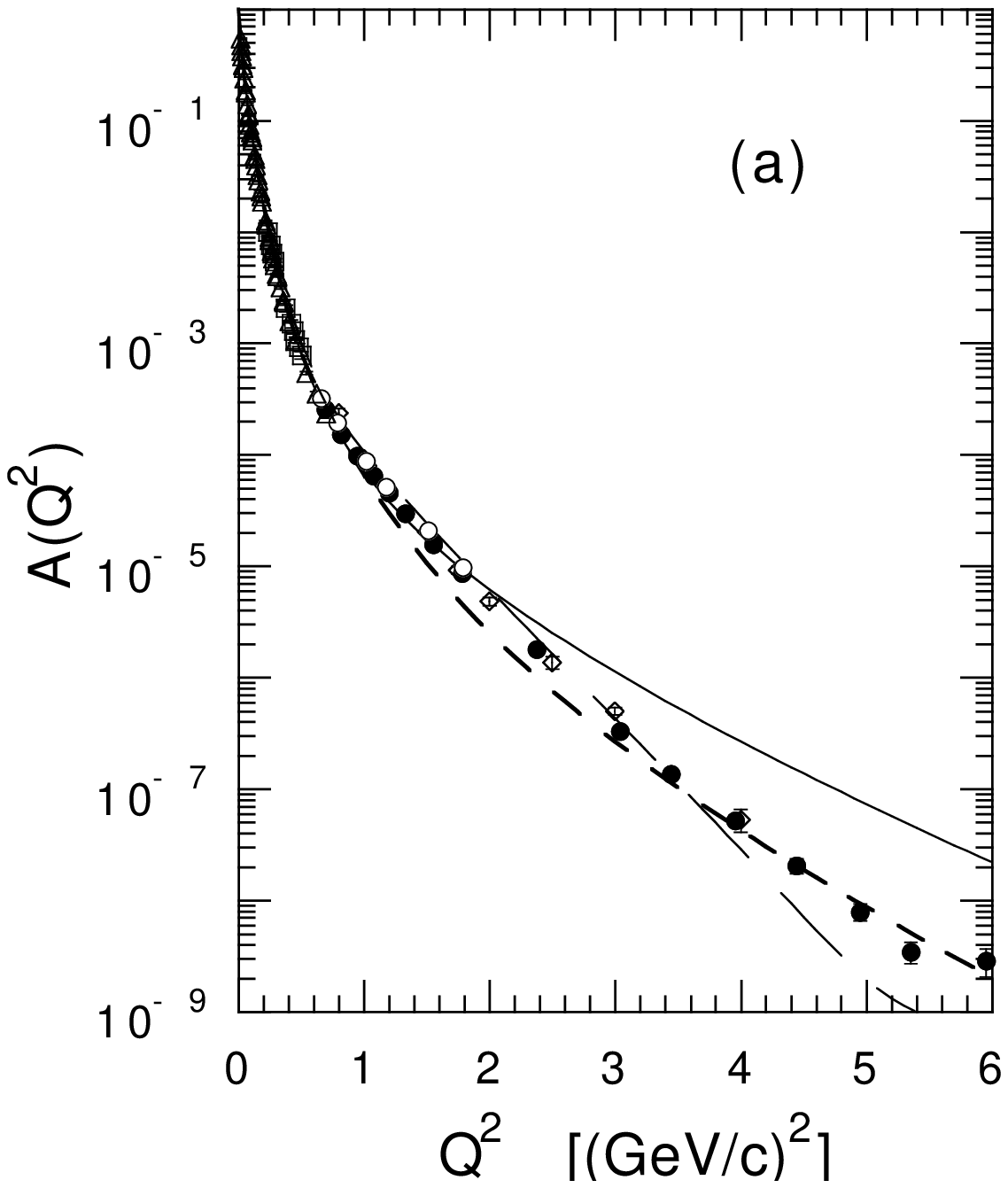}}
\   \ 
\parbox{6.4cm}{\includegraphics[width=6.4cm] {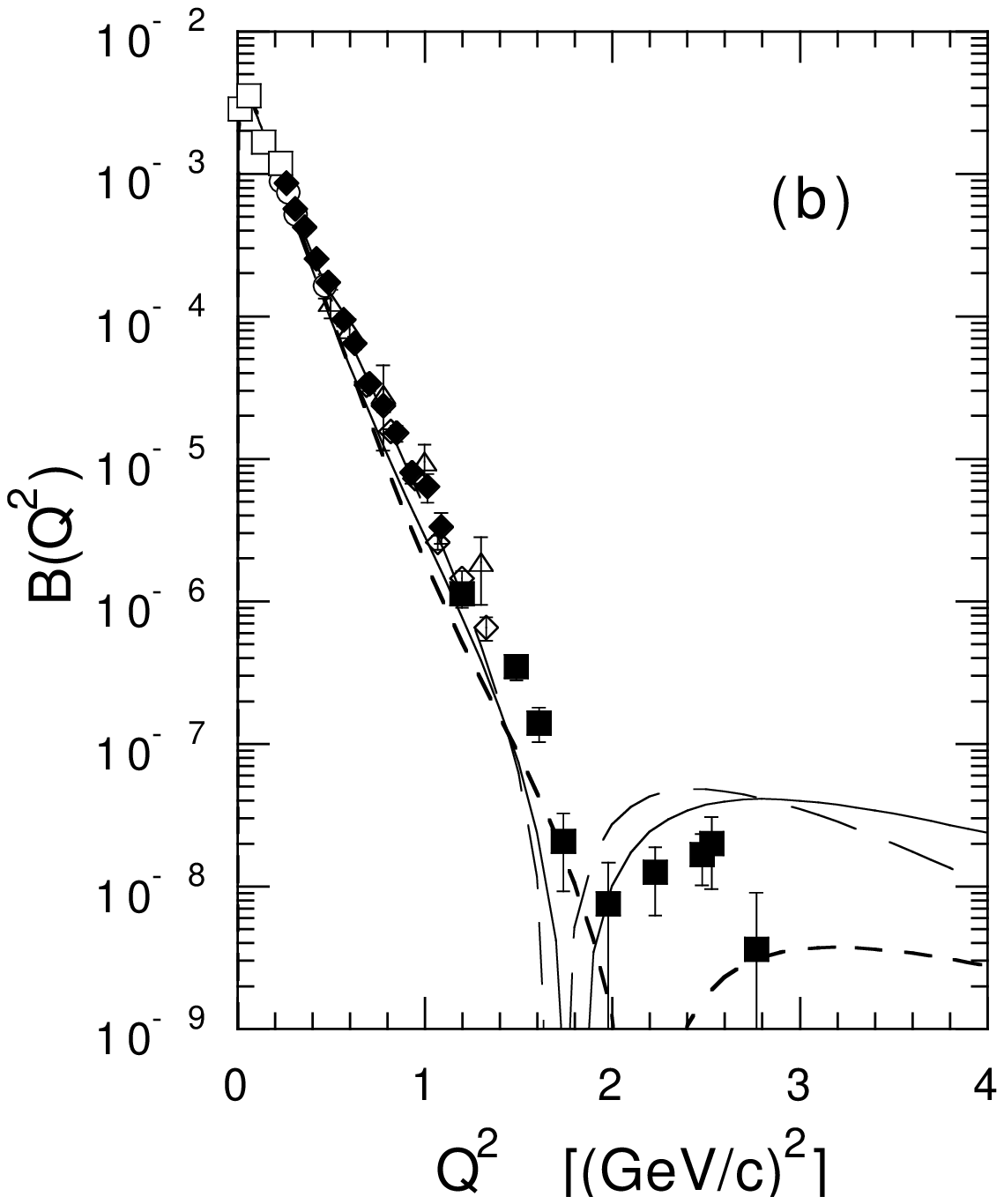}}

\caption{ 
$RSC$ $N-N$ interaction $+$  Gari-Kr\"{u}mpelmann Nucleon ff's.
 \cite{GK}. Solid line: LF full result with the
Poincar\'e covariant current operator, in the Breit frame where 
${\bf q}_\perp
=0$. 
 Dashed line: the same as the solid line, but the argument of the Nucleon ff's, 
 $(p_1' - p_1)^2$, is replaced by $ -Q^2$. Long-dashed line:
non relativistic result in the same Breit frame.(After Ref. \cite{LPS2K})}
 \label{H21}

\includegraphics[width=12.cm] {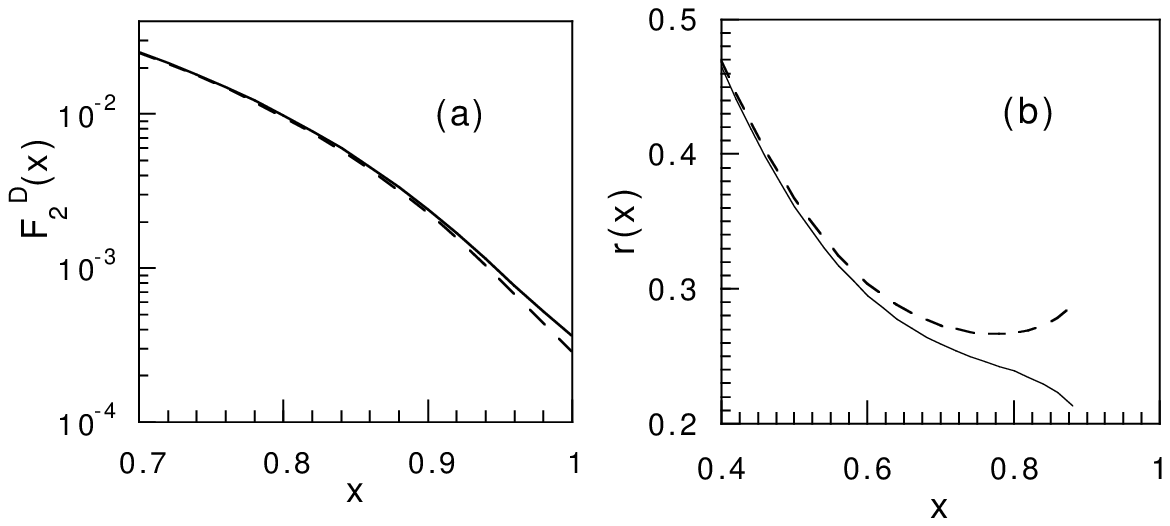}

\caption{Left panel: the $^2$H structure function, $F_2^D(x)$, vs. the Bjorken variable
$x$. Solid lines:  LF calculations, see Eq. (\ref{fLF}). 
Dashed lines:   standard approach calculations.
 Right panel: the
same as in the left panel but for the ratio $r(x)=F_2^n(x)/F_2^p(x)$. 
AV18  NN interaction and the model of Ref. \cite{Aubert} for the Nucleon structure functions  are adopted. 
(After Ref. \cite{PSCor}).}
\end{figure}

A very interesting topic, related indeed to both A=2 and A=3 nuclei, given the absence
of free neutron targets, is the extraction of the  ratio of Nucleon structure
functions, $r(x)=F_2^n(x)/F_2^p(x)$, in the Bjorken limit. This extraction has been
analyzed adopting a standard approach \cite{KPSS}, and here we would simply
recall that a LF approach could give remarkably different results. 
By using
the
Impulse Approximation, the Deuteron structure function can be expressed 
 through   a convolution of the Nucleon structure 
functions 
and the distribution probability, $f^{D}(z)$, to find inside $^2$H a
nucleon with LF momentum $z$.
In the usual approach, $f^D(z)$ is basically an adapted  Instant-form approach
with an off-mass-shell struck Nucleon, while 
within  a rigorous LF approach \cite{PSCor} it becomes
\be 
f^{D}_{LF}(z) = 
\int\nolimits d {\vec p} \; n^D( |{\vec p} |) \; \delta \left( z -  {\xi}
\frac {M_D} {M} \right)   
\label{fLF}\ee
where $n^D( |{\vec p} |)$ is the Nucleon momentum distribution in the Deuteron,
$M ( M_D )$ the Nucleon (Deuteron) mass, $ \xi = p^+/P^+=(\sqrt{M^2 +|\vec{p}|^2} +p_z)/2
\sqrt{M^2 +|\vec{p}|^2} $.

Following \cite{KPSS} one can  extract the ratio
$r(x)$ from the experimental data for the Deuteron 
structure
function and a suitable recurrence relation, where $f_D(x)$ has to be 
considered. In Fig. 2, one sees the interesting effect on the Deuteron 
structure function and on $r(x)$ (cf.  
  $x\to 1$), produced by  a LF $f^{D}(z)$, Eq.  (\ref{fLF}).

Trinucleon em observables, see Table 2 and Figs. 3-4, can be evaluated analogously  to $^2$H case, 
but with a very cumbersome angular coupling. In this preliminary calculation 
 ($S$, $P$ and $D$ waves only for observables in Table 2) the A=3 wave function \cite{KVR94} without Coulomb interaction and corresponding to 
  AV18 \cite{AV18} NN interaction has been used. The charge and magnetic 
  ff's are calculated from
 the matrix elements of a Poincar\'e covariant current as follows
$$
    F_{ch}^{ T_z}(Q^2) = {1 \over 2} ~ Tr[{\cal{I}}^+(T_z)] 
   \quad \quad \quad  F_{mag}^{ T_z}(Q^2) = - i { M \over Q} ~ Tr[{\sigma}_y ~
    {\cal{I}}_x(T_z)]    
$$ where $
     {\cal{I}}^{r}_{\sigma' \sigma }(T_z) \equiv 
    \langle \Psi_{{1 \over 2} \sigma'}^{{1 \over 2} T_z},
    {P}_f | ~ {\cal J}^{r} ~ |\Psi_{{1 \over 2} \sigma}^{{1 \over 2} T_z}, 
    {P}_i \rangle 
   $ with $r=+,1$ (cf Eq. (\ref{curlf})). 
\begin{table}\caption { Trinucleon magnetic moments and charge radii. Preliminary 
calculation with S+P+D waves:
${\cal P}_{S+S'}(Av18)\sim ~91.4$\%   
${\cal P}_{P}(Av18)\sim~0.07  $\% ${\cal P}_{D}(Av18)\sim~8.5  $\%. }
\begin{center} 
\begin{tabular} {|c||c| c|c| c|}
\hline
 Theory &  $\mu(^3\rm{He})$ & $\mu(^3\rm{H})$ & $r_{ch}(^3\rm{He})$  &$r_{ch}(^3\rm{H})$ \\
 \hline
NR(S) &-1.723(2) & 2.515(2) &1.841(3)  fm &1.798(3)  fm\\
\hline
LFD(S) & -1.7860(2) &  2.6034(2)&1.867(3)  fm & 1.821(3)  fm\\
\hline
NR(S+S') & -1.7093(2)  & 2.515(2)& 1.896(3) &1.726(3)  fm\\
\hline
LFD(S+S')  & -1.768(2)  & 2.600(2) & 1.919(3) &1.772(3)  fm\\
\hline
NR(S+S'+P+D) & -1.769(2) & 2.579(2) & 1.882(4)  fm & 
1.714(4)  fm\\
\hline
LFD(S+S'+P+D)  & -1.839(2)& 2.674(2) & 1.906(4) fm & 
1.754(4) fm\\
\hline
\hline
Exp. &  -2.1276 & 2.9789 & 1.959(30)fm & 1.755(86) fm\\
\hline
\end{tabular}
\end{center}
\end{table}
\begin{figure} 
\parbox{6.5cm}{\includegraphics[width=6.5cm,angle=0] {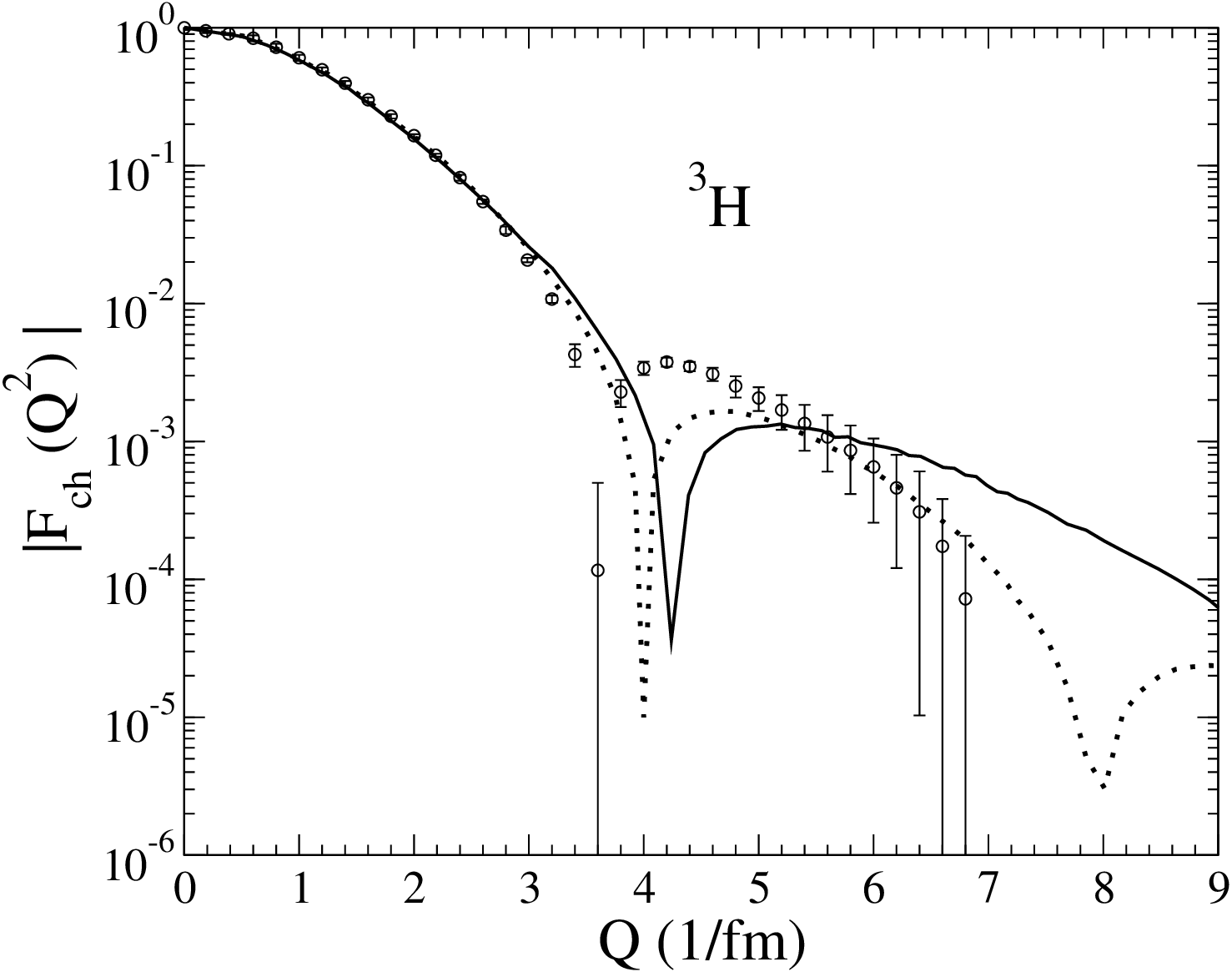}} 
\    \ 
\parbox{6.5cm}{\includegraphics[width=6.5cm,,angle=0] {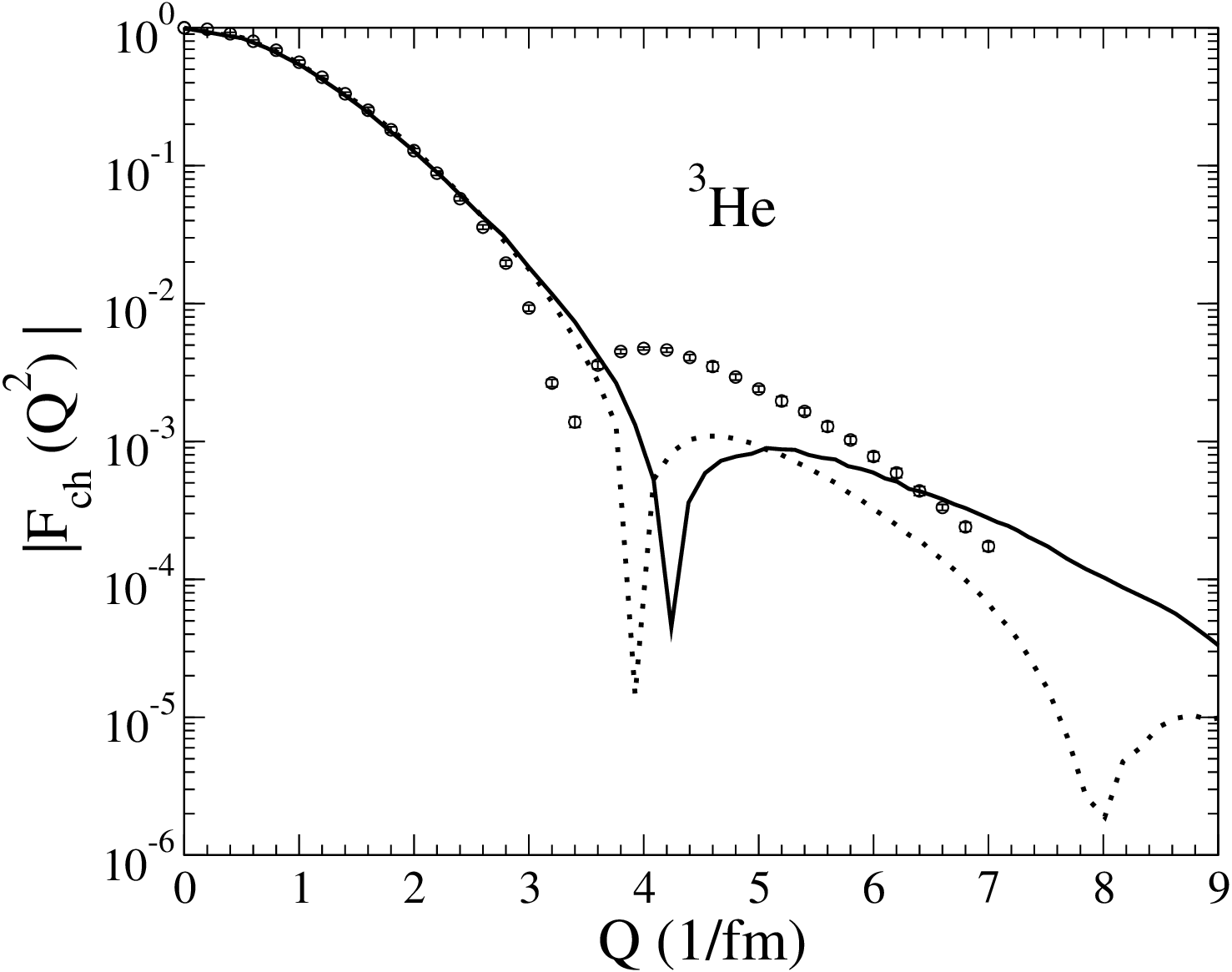}} 

\caption{Trinucleon charge ff vs $Q$, only S-wave. Left panel: $^3$H. 
Right panel:
$^3$He. Solid line: LFD calculations in a Breit frame where ${\bf q}_\perp=0$. 
Dotted line: non relativistic calculations in the same frame. 
AV18 trinucleon wave function \cite{KVR94} and  Gari-Kr\"umpelmann Nucleon 
ff's \cite{GK} have been adopted. 
Data from \cite{Sick}.}
 
\parbox{6.5cm}{\includegraphics[width=6.5cm,angle=0] {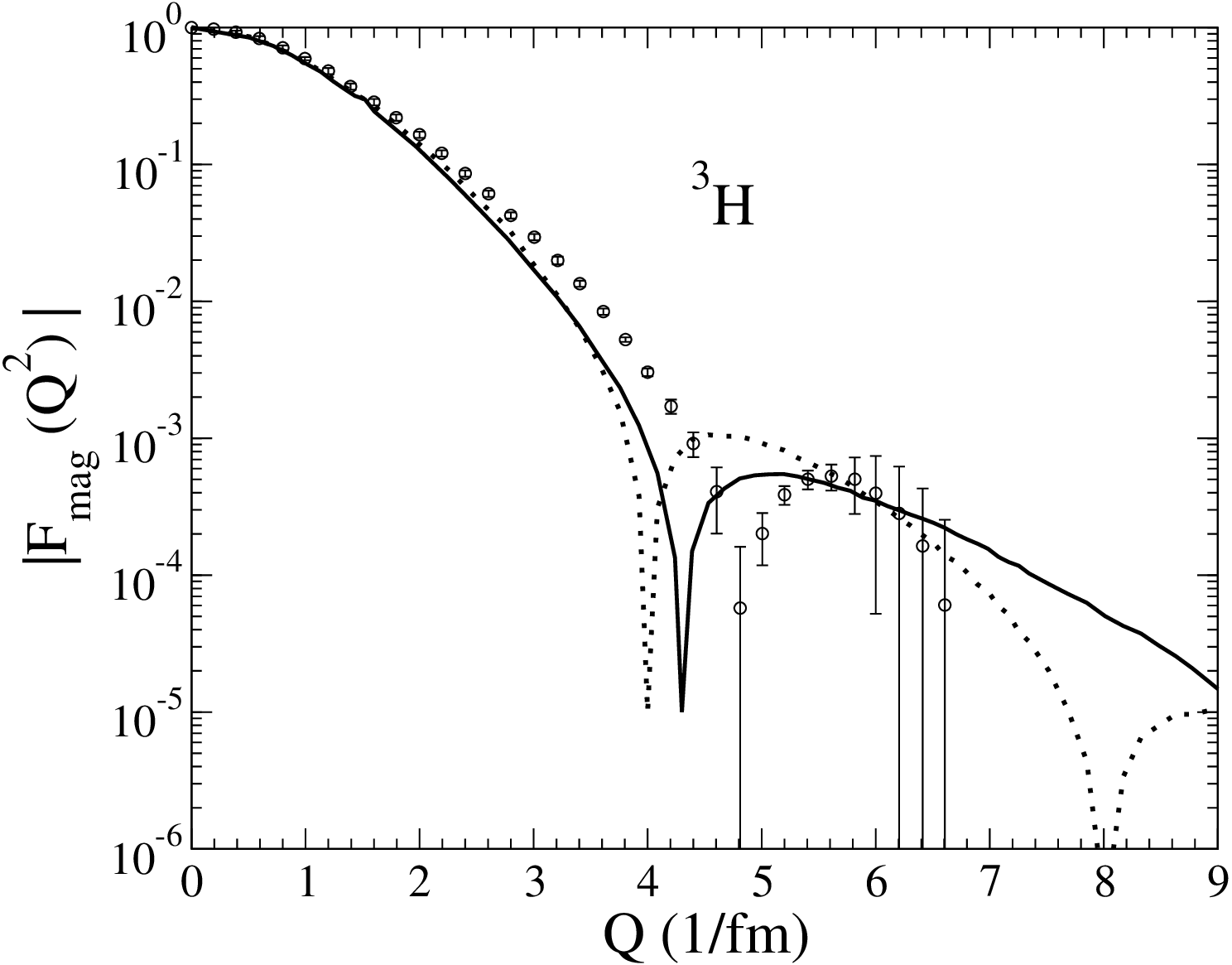}} 
\    \ 
\parbox{6.5cm}{\includegraphics[width=6.5cm,angle=0] {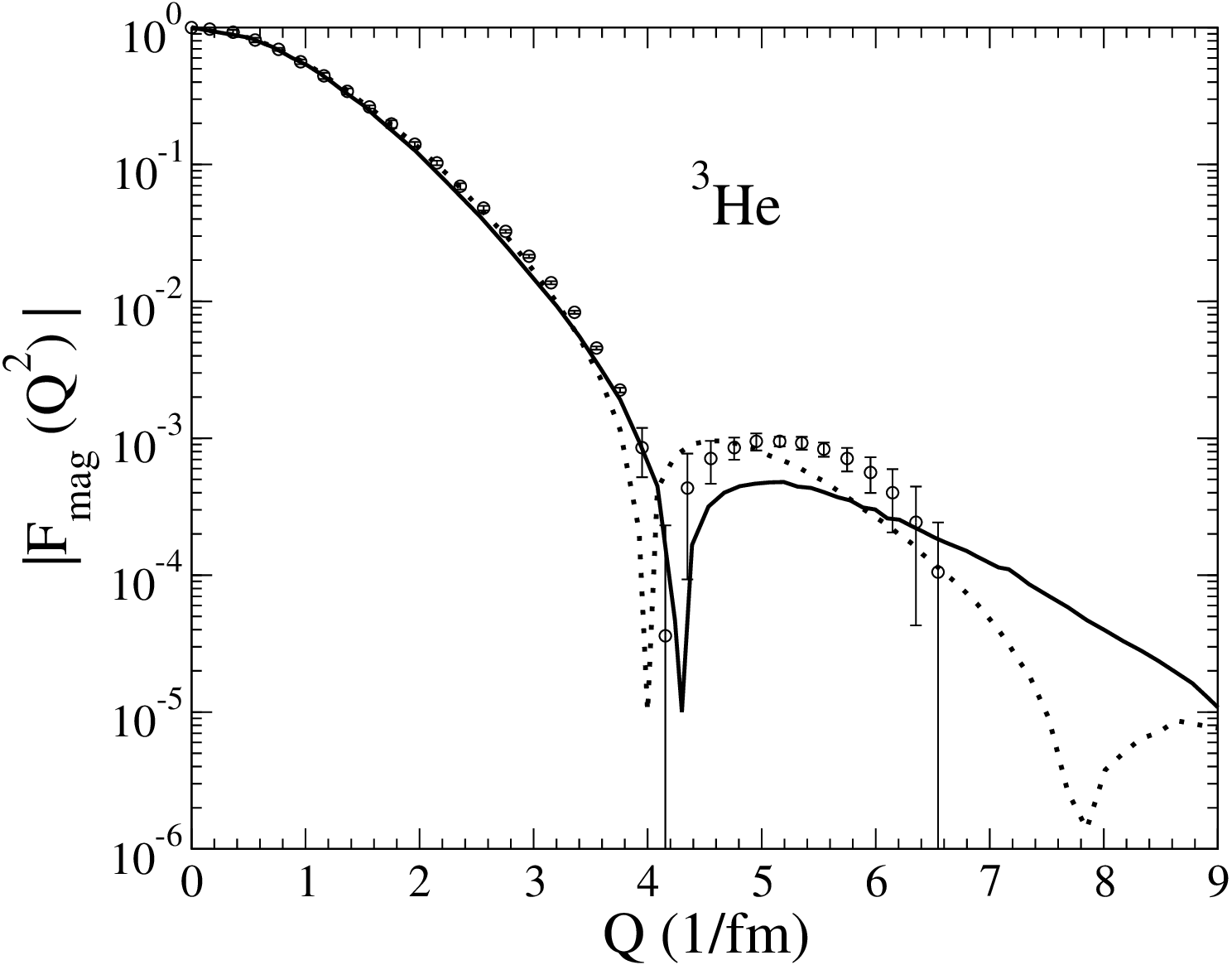}} 

\caption{The same as in Fig. 3 but for the trinucleon magnetic ff.}
\end{figure}

\section{Conclusions \& Perspectives}
 In order to construct a  {\em  standard model} for Few-Nucleon Systems
 it is necessary to consider relativistic effects. For a fixed number
 of constituents, one can be  a  Poincar\'e covariant approach
 by  adopting  a
 Light-Front RHD and  the Bakamjian-Thomas procedure. Within such an approach
 and  the em
 current operator suggested in  Ref. \cite{LPS98}, 
 the em observables
 of the A=3 nuclei have been calculated for the first time ($S$, $P$ and $D$ waves for
 the observables in Table 2 and $S$-wave only for ff's in Figs. 3-4).
 The few \% effect for em observables at $Q^2=0$, 
 in the correct direction, is very encouraging. Moreover,
 the sizable effect at high $Q^2$ indicates the essential role played by relativity
 for analyzing the em ff's in the region of few GeV's.
 A full calculation, with a systematic analysis of the pair contribution 
 (Z-diagram) and 
  current operators fulfilling the Ward-Takahashi Identity, will be presented
  elsewhere.

\begin{acknowledge}
We gratefully thank  Alejandro Kievsky for providing us the trinucleon
wave function corresponding to  the  AV18 NN interaction.
\end{acknowledge}

\end{document}